%% file: main-arxiv.tex
\documentclass[acmlarge,nonacm]{acmart}
\makeatletter
\AtBeginDocument{
  \fancypagestyle{firstpagestyle}{
    \fancyhead{}%
    \fancyfoot[C]{}%
  }
  \fancyhf{}
  \fancyhead[LO]{\@headfootfont\shorttitle}%
  \fancyhead[RE]{\@headfootfont\@shortauthors}%
  \fancyhead[LE]{}%
  \fancyhead[RO]{}%
  \fancyfoot[C]{}%
}
\makeatother

\AtBeginDocument{\settopmatter{printacmref=false}}

\copyrightyear{2026}
\acmYear{2026}
\setcopyright{none}

\usepackage{cleveref}
\usepackage{multirow}
\usepackage{booktabs}
\usepackage{array}
\usepackage{tabularx}
\usepackage{xurl}
\usepackage{microtype}
\usepackage{balance}

\begin{document}

\title{Societal Frameworks Can Improve LLM Alignment}

\author{Karolina Sta\'nczak}
\thanks{Correspondence to: karolinaewa.stanczak@ai.ethz.ch}
\orcid{0000-0001-7326-9594}
\affiliation{%
  \institution{ETH Zurich}
  \city{Zurich}
  \country{Switzerland}
}
\affiliation{%
  \institution{Mila, McGill University}
  \city{Montreal}
  \country{Canada}
}

\author{Nicholas Meade}
\affiliation{%
  \institution{Mila, McGill University}
  \city{Montreal}
  \country{Canada}
}

\author{Mehar Bhatia}
\affiliation{%
  \institution{Mila, McGill University}
  \city{Montreal}
  \country{Canada}
}

\author{Hattie Zhou}
\authornote{Work done by HZ prior to joining Anthropic.}
\affiliation{%
  \institution{Mila, Universit\'e de Montr\'eal}
  \city{Montreal}
  \country{Canada}
}
\affiliation{%
  \institution{Anthropic}
  \city{San Francisco}
  \country{USA}
}

\author{Konstantin B\"ottinger}
\affiliation{%
  \institution{Fraunhofer AISEC}
  \city{Munich}
  \country{Germany}
}

\author{Jeremy Barnes}
\affiliation{%
  \institution{ServiceNow}
  \city{Montreal}
  \country{Canada}
}

\author{Jason Stanley}
\affiliation{%
  \institution{ServiceNow}
  \city{Montreal}
  \country{Canada}
}

\author{Jessica Montgomery}
\affiliation{%
  \institution{University of Cambridge}
  \city{Cambridge}
  \country{UK}
}

\author{Richard Zemel}
\affiliation{%
  \institution{Columbia University}
  \city{New York}
  \country{USA}
}

\author{Nicolas Papernot}
\affiliation{%
  \institution{University of Toronto, Google DeepMind}
  \city{Toronto}
  \country{Canada}
}

\author{Nicolas Chapados}
\affiliation{%
  \institution{Mila, ServiceNow}
  \city{Montreal}
  \country{Canada}
}

\author{Denis Therien}
\affiliation{%
  \institution{McGill University, ServiceNow}
  \city{Montreal}
  \country{Canada}
}

\author{Timothy P. Lillicrap}
\affiliation{%
  \institution{Google DeepMind}
  \city{London}
  \country{UK}
}

\author{Ana Marasovi\'c}
\affiliation{%
  \institution{University of Utah}
  \city{Salt Lake City}
  \country{USA}
}

\author{Sylvie Delacroix}
\affiliation{%
  \institution{King's College London}
  \city{London}
  \country{UK}
}

\author{Gillian K. Hadfield}
\affiliation{%
  \institution{Johns Hopkins University}
  \city{Baltimore}
  \country{USA}
}

\author{Siva Reddy}
\affiliation{%
  \institution{Mila, McGill University, ServiceNow}
  \city{Montreal}
  \country{Canada}
}
\renewcommand{\shortauthors}{Sta\'nczak et al.}

\begin{abstract}
Recent progress in large language models (LLMs) has focused on producing responses that meet human expectations and align with shared values\,---\,a process coined \textit{alignment}. 
However, aligning LLMs remains challenging due to the inherent disconnect between the complexity of human values and the narrow nature of the technological approaches designed to address them. 
Current alignment methods often lead to misspecified objectives, reflecting the broader issue of \textit{incomplete contracts}, the impracticality of specifying a contract between a model developer and the model that accounts for every scenario in LLM alignment. 
In this paper, we argue that improving LLM alignment requires incorporating insights from societal alignment frameworks, including social, economic, and contractual alignment, and discuss potential solutions drawn from these domains. 
Given the role of uncertainty within societal alignment frameworks, we then investigate how it manifests in LLM alignment.
It is this pervasive uncertainty that necessitates our alternative view on LLM alignment, framing the under-specified nature of its objectives as an opportunity rather than perfect their specification. Beyond technical improvements in LLM alignment, we discuss the need for participatory alignment interface designs.
\end{abstract}




\maketitle

\input{sections/01_intro}

\input{sections/02_alignment}

\input{sections/03_contract}

\input{sections/04_societal_frameworks}

\input{sections/05_uncertainty}

\input{sections/06_alternative}

\input{sections/07_related_work}

\input{sections/08_conclusion}

\section*{Generative AI Usage Statement}

In accordance with the ACM Policy on Authorship regarding the use of generative AI, the authors disclose that generative AI tools (specifically Gemini 3 Pro) were used in the preparation of this manuscript. The use of these tools was strictly limited to: (1) assisting with grammar, syntax, and the polishing of author-written text for fluency; and (2) assisting with the technical formatting of LaTeX.
These tools were not used to generate new research content, conceptual ideas, or citations. All references were manually collected. The authors edited all outputs and remain fully accountable for this work.\looseness=-1

\begin{acks}
This paper originated from the Bellairs Invitational Workshop on Contemporary, Foreseeable, and Catastrophic Risks of Large Language Models in April 2024. We thank all workshop participants for their valuable discussions and contributions. Karolina Sta\'nczak was supported by the Mila P2v5 grant, the Mila-Samsung grant, and by the ETH AI Center postdoctoral fellowship.
\end{acks}

\bibliographystyle{ACM-Reference-Format}
\bibliography{sample-base,custom}

\appendix

\end{document}

%% file: sections/01_intro.tex
\section{Introduction}

As large language models (LLMs) advance to unprecedented levels of proficiency in generating human-like language, aligning their behavior with human values has become a critical challenge to ensuring their usability in real-world applications \citep{leike2018scalableagentalignmentreward,gabriel2020artificial,NEURIPS2022_b1efde53,shen2023largelanguagemodelalignment}.
This alignment encompasses both \textit{explicit} values, such as following instructions and being helpful, and \textit{implicit} values, such as remaining truthful and avoiding biased or otherwise harmful outputs \citep{askell2021generallanguageassistantlaboratory}.
In fact, the rise of LLM-based chat assistants has largely been driven by their ability to follow instructions and engage in open-ended dialogue, demonstrating the importance of alignment, enabled by algorithms such as reinforcement learning from human feedback (RLHF; \cite{NEURIPS2022_b1efde53,ziegler2020finetuninglanguagemodelshuman}).

Despite these advancements, aligning LLMs with human values remains a formidable challenge \citep{wei2023jailbroken,williams2024targetedmanipulationdeceptionoptimizing,greenblatt2024alignmentfakinglargelanguage}. 
This difficulty primarily stems from the fundamental gap between the intricacies of human values and the often narrow technological solutions \citep{hadfieldmanell2019}. 
Current LLM alignment methods, such as RLHF, often result in misspecified alignment objectives, where reward functions reflect human values only within designer (or annotators) provided scenarios, a finite set among an infinite set of values, failing to generalize in unforeseen contexts \citep{amodei2016concreteproblemsaisafety,hadfieldmanell2019,10.1145/3375627.3375851,NEURIPS2021_c26820b8,10.5555/3600270.3600957}. While developers acknowledge the problem of misspecification \citep{leike2018scalableagentalignmentreward, shen2023largelanguagemodelalignment,NEURIPS2022_b1efde53}, the root causes of this issue have been largely overlooked.

\begin{figure*}[ht]
    \centering
    \includegraphics[trim={0 12.5cm 0 0},clip,width=\textwidth]{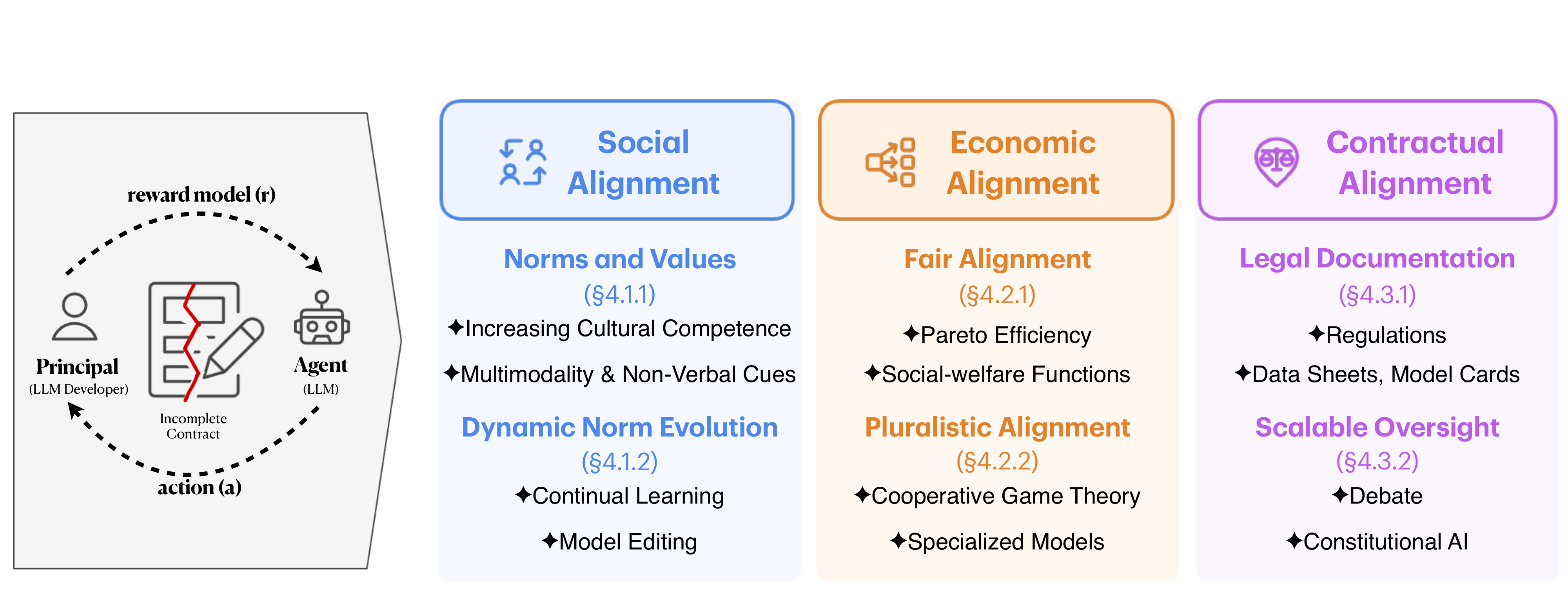}
    \caption{We view human-LLM interactions as a \textit{principal}-\textit{agent} framework, where a \textit{principal} (a system designer) incentivizes an \textit{agent} (an LLM) to take an action $a$ by offering a reward $r$. This framework assumes that the agent's action is driven by its reward function, forming a pair $(a,r)$ that serves as a \emph{contract} between the agent and the principal. However, this contract is incomplete. To address this incompleteness, we explore societal alignment mechanisms of social, economic, and contractual alignment as guiding principles for LLM alignment in the incomplete contracting environment.\looseness=-1}
    \Description{Hand-made diagram illustrating the concepts of societal alignment in relation to large language models.}
  \label{fig:your_label}
    \label{fig:intro}
\end{figure*}

To better understand this misalignment, we frame LLM alignment within a \emph{principal-agent}\footnote{We use `agent' in the contract theory sense, referring to an entity acting on behalf of a principal, rather than the broader AI notion of autonomous systems.} framework \citep{f376b2cf-a8b4-3622-927b-3396b646b772}, a well-established paradigm in economic theory. As shown in \Cref{fig:intro}, in this framework, the LLM acts as the agent and the model developer (or user) serves as the principal. We define a \emph{contract} as a pair: an action taken by the agent and the corresponding reward assigned by the principal. For example, a contract in LLM training could reward the model for generating responses that follow factual accuracy constraints while penalizing hallucinated outputs. The principal is able to steer the agent's behavior toward intended objectives with an appropriate reward. In an ideal scenario, a complete contract would perfectly align the agent's actions with the principal's objectives in all possible states of the world.\looseness=-1

However, designing a fully specified contract that anticipates every possible scenario in model training is infeasible \citep{hadfieldmanell2019,zhuang2020consequences}. In LLM alignment, this challenge is reflected in the reward function, which is derived from explicitly elicited values or implicitly implied values in the form of human preferences.
Yet, quantifying complex and often diverging human values is difficult \citep{leike2018scalableagentalignmentreward,feffer2023moralmachinetyrannymajority}, and capturing them effectively incurs high annotation costs \citep{klingefjord2024humanvaluesalignai}. Aggregating these values into a unified reward signal is nontrivial \citep{Kemmer_Yoo_Escobedo_Maciejewski_2020,ilvento:LIPIcs.FORC.2020.2}.

 
These alignment challenges are not unique to LLMs. In fact, they echo broader alignment problems that humans encounter daily due to incomplete contracts.
Institutions such as society, economy, and law enable us to thrive despite incompleteness.
In this position piece, \textbf{we advocate for leveraging insights from societal alignment frameworks to guide the development of LLM alignment within incomplete contracting environments}. 
Drawing on principles from social alignment (\Cref{sec:social}), economic alignment (\Cref{sec:economic}), and contractual alignment (\Cref{sec:legal}), we propose solutions to guide behavior in incomplete contracting environments, much like they have for human societies (see \Cref{fig:intro}). 
By design, we strategically refrain from prescribing technical implementation details, as these research directions each demand collective and dedicated effort beyond the scope of this paper.

We then explore how uncertainty, inherent in incomplete contracting environments \citep{seita1984uncertainty}, manifests in LLM alignment (\Cref{sec:uncertainty}). For instance, an LLM analyzing patient symptoms without a full medical history must navigate this ambiguity to avoid user confusion or misinterpretation.
We contend this pervasive uncertainty is not merely a technical challenge but a fundamental characteristic demonstrating that any single, pre-defined stage of alignment will prove insufficient. It is precisely this irreducible uncertainty that necessitates the next stage of alignment, which we detail in our alternative view (\Cref{sec:alternative}).
This alternative view does not oppose existing societal frameworks; rather, it highlights their indispensability while showing why they alone are not enough. Accordingly, we frame an LLM's under-specified objectives not as a flaw to be resolved solely by technological alignment, but as an opportunity for participatory alignment that actively engages diverse stakeholders in LLM alignment.\looseness=-1

%% file: sections/02_alignment.tex
\section{Contemporary Approach to LLM Alignment}
\label{sec:alignment}

Aligning LLMs with human values is commonly understood as training them to act in accordance with user intentions \citep{leike2018scalableagentalignmentreward}.
The objective of LLM alignment is often conceptualized as fulfilling three core qualities, often referred to as the ``3H'' framework: honesty (regarding their capabilities, internal states, and knowledge), helpfulness (in performing requested tasks or answering questions within safe bounds), and harmlessness (encompassing both the refusal to fulfill harmful requests and the avoidance of generating harmful content) \citep{askell2021generallanguageassistantlaboratory,Bai2022TrainingAH}.\looseness=-1 

A prominent approach to achieve this alignment is through a preference-based approach like RLHF. The RLHF pipeline usually includes three stages: supervised fine-tuning (SFT), preference sampling and reward model training \citep{NIPS2017_d5e2c0ad,steinnon2020learning}, and reinforcement learning fine-tuning either using proximal policy optimization (PPO; \citep{schulman2017proximalpolicyoptimizationalgorithms}), or directly through policy optimization (DPO; \citep{rafailov2023direct}). The process usually starts with a generic pre-trained language model, which undergoes supervised learning on a high-quality dataset for specific downstream tasks. In this paper, we focus on the implications of the reward modeling stage due to its connection to an incomplete contract, which we will lay out in \Cref{sec:contract}.\looseness=-1

\subsection{Reward modeling from human preference.}
In the reward modeling stage, for a given input prompt $x$, the SFT model generates paired outputs, ${y_0, y_1} \in \mathcal{Y} \times \mathcal{Y}$, where $\mathcal{Y}$ denotes the set of all possible outputs that the model can generate in response to a given input. Human evaluators then select their preferred response, $y \in {y_0, y_1}$, providing data that guides the alignment process \citep{NIPS2017_d5e2c0ad,steinnon2020learning}. 
Human preferences are modeled probabilistically using frameworks like the Bradley-Terry model \citep{19ff28b9-64f9-3656-ba40-08326a05748e}. The preference probability for one response over another is expressed as
\begin{equation}
p(y_1 \succ y_2 \mid x) = \frac{\exp(r(x, y_1))}{\exp(r(x, y_1)) + \exp(r(x, y_2))},
\end{equation}
where $r(x, y)$ is a latent reward function approximated by a parametric reward model, $r_\phi(x, y)$. Using a dataset of comparisons $\mathcal{D}$, the reward model is trained by minimizing the negative log-likelihood
\begin{equation}
\mathcal{L}_R(r_\phi, \mathcal{D}) = -\mathbb{E}_{(x, y_w, y_l) \sim \mathcal{D}} \Big[\log \sigma \big( r_\phi(x, y_w) - r_\phi(x, y_l) \big)\Big],
\end{equation}
where $\sigma$ is the logistic function and $y_w$ and $y_l$ denote the preferred and dispreferred completions among $(y_1, y_2)$.

%% file: sections/03_contract.tex
\section{LLM Alignment as a Contract}
\label{sec:contract}

In the following, we formalize LLM alignment through the lens of contract theory \citep{RePEc:mtp:titles:0262025760,echenique2023online}, a subfield of economics that studies how agreements are designed under conditions of incomplete information.
We describe human-LLM interactions as a \textit{principal-agent} relationship, where a \textit{principal} (e.g., the user, system designer, or a company) seeks to incentivize an \textit{agent} (an LLM) to act in a desired manner \citep{garen1994} (see \Cref{fig:intro}). 
This framework provides a way to conceptualize how the principal tries to align the agent's behavior with their objectives, using the agent's action and its reward function as a \textit{contract}. In this section, we explore the contract formalization (\Cref{sec:contract-formal}) and how the incompleteness of this contract (\Cref{sec:contract-incomplete}) directly leads to misalignment (\Cref{sec:contract-misalignment}) in the context of LLM alignment.\looseness=-1

\subsection{Contract Formalization}
\label{sec:contract-formal}
Following \citet{echenique2023online}, we define a contract as a pair $(a, r)$, where $a \in \mathcal{A}$  represents an action of an agent and $r:(\mathcal{X} \times \mathcal{Y}) \to \mathbb{R}$ is a reward function.\footnote{Here we loosely refer $a$ to mean one action or a series of actions that lead to an LLM output.}
The function $r$ determines the agent's reward based on the observed input-output pair $(x, y)$. In the context of a user-LLM interaction, an input $x \in \mathcal{X}$ corresponds to a user prompt, and output $y \in \mathcal{Y}$ is the LLM-generated response. A contract might be, for instance, a positive reward if the model avoids hate speech in the output. Here, the reward function would be trained on prompt-response pairs, awarding higher scores to responses that do not contain hate speech.\looseness=-1 

The framework is initiated, for instance, when a user, acting as the principal, initiates the interaction by prompting an LLM, thus implicitly proposing a contract. 
The LLM, acting as an agent, then implicitly either accepts or rejects this contract. Rejection of the contract manifests in the LLM not converging towards the desired output, which is a generated response without hate speech.
Upon implicitly accepting the contract, the LLM conducts an action $a$, which can be viewed as a probability distribution over all possible model outputs that satisfy the contract. We note that the user does not directly observe the LLM's internal decision of its action but only the output $y$. Consequently, the agent is rewarded according to the agreed-upon reward function, $r(x,y)$, implemented as a reward signal during the training phase. The principal experiences the utility derived from the output $y$; that is, the user benefits from the generated response but also suffers if the model behaves adversarially. This is illustrated when, despite a contract penalizing hate speech, the LLM generates responses that subtly convey harmful biases.\looseness=-1

\subsection{The Challenge of Incomplete Contracting in AI}
\label{sec:contract-incomplete}

Although the specific implications of incomplete contracting for LLM alignment remain underexplored, the concept has been studied in the broader context of AI alignment \citep{hadfieldmanell2019}.
In theory, alignment between the principal and the agent theoretically requires a \textit{complete contract} \citep{williamson1975markets,hadfieldmanell2019}. A complete contract would perfectly align the principal's objectives with the agent's behavior in all possible states of the world. This requires that action $a$ and reward function $r(x,y)$ be optimally defined for all input-output pairs. However, achieving complete contracts is practically infeasible for AI systems, rendering incomplete contracting unavoidable \citep{hadfieldmanell2019}. This is primarily due to the fact that machine learning systems inherently operate with underspecified objectives \citep{10.5555/3586589.3586815}, which stems from the practical difficulty in defining a reward function $r(x,y)$ that fully captures the complexities of the desired behavior.\looseness=-1

The difficulty in specifying such a complete reward function arises from several issues.
First, a key challenge for AI alignment generally, real-world applications are too complex to generate all possibilities, hindering the specification of every possible $(a,r)$ pair \citep{openai_faulty_reward_functions}. The space of possible outcomes, denoted by $\mathcal{Y}$ in the formalization is not tractable. This mirrors the challenge of LLM in generating outputs for new input it might receive during inference. The challenge extends beyond the practical limitations of fully specifying objectives. 
Second, and particularly relevant for LLMs, even beyond these practical limitations, the challenge of translating complex human values into reward functions remains. Ambiguities and gaps in defining the desired action contribute to unintended and often undesirable outcomes.\looseness=-1

\subsection{Misalignment due to an Incomplete Contract} 
\label{sec:contract-misalignment}

We frame LLM alignment as a challenge of incomplete contracting, which leads to misalignment. In the context of LLMs, this misalignment occurs when the reward function, $r(x,y)$ is underspecified, and thus might incentivize outputs that diverge from the users's true objectives.\looseness=-1

A common outcome of reward misspecification is \textit{reward hacking}, where an agent optimizes for the reward itself rather than the intended behavior. 
For example, LLMs may exploit gaps in the specifications, such as in the ``jailbreaking'' phenomenon. Here, carefully crafted prompts elicit harmful responses by bypassing weak guardrails because their reward function is not specific enough, allowing the model to optimize without complying with safety requirements \citep{chao2024jailbreakingblackboxlarge,zou2023universaltransferableadversarialattacks}. 
Another example of reward hacking is an LLM trained to generate ``helpful'' responses might learn to produce lengthy and verbose answers to prompts, as this might result in a higher score from the reward function even if it is not actually helpful to the user \citep{saito2023verbosity}. 
A related issue, ``fake alignment,'' occurs where the agents superficially comply with the training objective without adopting the intended internal goals \citep{greenblatt2024alignmentfakinglargelanguage}.
Another challenge is the \textit{inherent context dependence} of reward functions, which need to adapt appropriately to evolving contexts. A contract might specify desired behavior in a narrow scenario, but leave ambiguities for broader applications \citep{NIPS2017_32fdab65}. For example, a contract that stipulates ``no harmful bias'' is inherently underspecified since the definitions of ``harmful'' and ``bias'' are context-dependent. 
Because a complete and exhaustive specification of desired behavior is fundamentally impossible, LLMs, much like human agents, require external frameworks to resolve the inherent ambiguity of incomplete contracts. Just as human society relies on implicit social norms, economic incentives, and legal principles to navigate underspecified contracts, we argue that AI systems must similarly be grounded in these societal frameworks to ensure robust alignment.

%% file: sections/04_societal_frameworks.tex
\section{Societal Alignment Frameworks}
\label{sec:societal-frameworks}

\begin{table*}[ht]
\small
\centering
\caption{Overview of Societal Alignment Frameworks. A breakdown of relevance, research directions, and frameworks for social, economic, and contractual alignment.}
\label{tab:alignment-frameworks}
\begin{tabular}{>{\centering\arraybackslash}m{2.2cm} >{\raggedright\arraybackslash}m{2cm} >{\raggedright\arraybackslash}m{3.3cm} >{\raggedright\arraybackslash}m{3.3cm} >{\raggedright\arraybackslash}m{3.3cm}}
\toprule
\textbf{Framework} & \textbf{Dimension} & \textbf{Relevance} & \textbf{Societal Frameworks} & \textbf{Implementations} \\ \midrule

\multirow{8}{2.3cm}{\centering\textbf{Social \\ Alignment}} & \textbf{Normative \newline Competence} & Reducing cultural bias; Explicitly defining latent norms & Normative systems \citep{Schutz1976}, cooperative intelligence \citep{mercier2018enigma} & Social norms corpora \citep{ziems-etal-2022-moral,zhan-etal-2024-renovi}, Cultural teaming \citep{chiu2024culturalteaming} \\ \cmidrule{2-5}
& \textbf{Multimodal \newline Cues} & Closing the ``normative void'' in human-AI interaction & Evolutionary Psychology: Non-verbal communication cues \citep{doi:10.1098/rstb.2013.0302} & Visual/auditory signal integration for intent recognition \cite{liu-etal-2021-visually,nayak-etal-2024-benchmarking} \\ \cmidrule{2-5}
& \textbf{Dynamic \newline Adaptation} & Preventing signals from outdated training data & Sociology: Fluidity of social norms and Dynamic Equilibria \citep{doi:10.1073/pnas.1817095116} & Model editing \citep{pmlr-v162-mitchell22a}, Continual learning for bias mitigation \\ \midrule

\multirow{6}{2.3cm}{\centering\textbf{Economic \\ Alignment}} & \textbf{Fair \newline Allocation} & Group-level equity; Balancing individual vs. collective utility & Welfare Economics: Social welfare functions \citep{arrow1951extension, DASPREMONT2002459} & Pareto-optimal preference learning \citep{boldi2024paretooptimallearningpreferenceshidden}, welfare-centric RLHF \citep{cousins2024welfare} \\ \cmidrule{2-5}
& \textbf{Pluralistic \newline Values} & Navigating conflicting preferences across diverse stakeholders & Game Theory: Cooperative games for resource/value allocation \citep{10.5555/2132771} & Pluralistic alignment \citep{SorensenMFGMRYJ24}, Few-shot learning for diverse perspectives value profiles \cite{sorensen-etal-2025-value} \\ \midrule

\multirow{5}{2.3cm}{\centering\textbf{Contractual \\ Alignment}} & \textbf{External \newline Oversight} & Standardized documentation and legal compliance & Contract Law: Incomplete contracting \citep{maskin1999unforseen, tirole1999incomplete} & Model Cards \citep{10.1145/3287560.3287596}, datasheets \citep{10.1145/3458723}, ethics guidelines (EU Commission) \\ \cmidrule{2-5}
& \textbf{Internal \newline Governance} & Embedding principles into model reasoning chains & Political Science: Democratic institutionalized processes \citep{shen2023largelanguagemodelalignment} & Constitutional AI \citep{bai2022constitutionalaiharmlessnessai}, AI safety debate \citep{irving2018aisafetydebate} \\ \bottomrule
\end{tabular}
\end{table*}

We present societal alignment frameworks that can provide guidelines for LLM alignment in an incomplete contracting environment. \Cref{tab:alignment-frameworks} provides an overview of these frameworks, their discussed dimensions, and their relevance to current research on LLM alignment. While we touch upon how their insights might inform the development of LLMs, we acknowledge that translating these frameworks into robust and scalable methods defines a new research direction. 
In the following, we discuss the alignment mechanisms of social theory (\Cref{sec:social}), economic theory (\Cref{sec:economic}), and contractual theory (\Cref{sec:legal}), and explore potential solutions and associated technical considerations for improving current LLM alignment approaches.\looseness=-1

\subsection{Social Alignment}
\label{sec:social}

Human communication relies on a complex, largely implicit set of norms, values, and cues that guide individuals in interpreting each other's intentions and the world around them \citep{10.1093/acprof:oso/9780190622046.001.0001}. 
However, this process is inherently ambiguous, as much of the meaning is conveyed implicitly rather than explicitly stated. Nonetheless, humans possess a unique ability called \textit{normative competence}, which allows them to understand and judge whether certain behaviors are appropriate or inappropriate in a given context \citep{Schutz1976}. This capability is often ingrained in cultures across the world \citep{hershcovich-etal-2022-challenges, doi:10.1177/0956797615586188,10.3389/fpsyg.2018.00849}, shaping shared understanding that facilitates communication and fosters mutual understanding \citep{mercier2018enigma}. 
A similar challenge arises in user-LLM interactions, where the absence of shared norms and values can result in misaligned outputs.
For example, an LLM providing evening activity recommendations without accounting for cultural context might suggest visiting a bar or consuming alcohol in a region where such activities are prohibited or socially unacceptable, leading to responses that fail to align with local norms. 

This gap highlights the need of what recent scholarship describes as ``thick'' conceptual representations \cite{tan2025beyond} and socially grounded contextual analysis for AI value alignment \cite{nelson2023thick, 10.1215/08992363-10742593}. In particular, \citet{kommers2025meaningmetricusingllms} argue that LLMs can be used to make cultural context, and thus human meaning, legible at an unprecedented scale within AI-based socio-technical systems. Internalizing these ``thick'' societal norms into LLMs could equip them with mechanisms to interpret and dynamically adapt to human normative systems \citep{10.5555/2447556.2447672}, much like these aid alignment within human interactions \citep{10.1093/acprof:oso/9780190622046.001.0001}.

\subsubsection{Instilling norms and values} 

While norms constitute context-dependent behavioral rules that individuals follow, values represent broader ideals representing overarching goals and aspirations, shaping what individuals strive for \citep{matsumoto}. As a fundamental tool of cooperative intelligence, language plays a crucial role in expressing and reinforcing both norms and values. 
These can be instilled during LLM alignment in several ways.\looseness=-1 

LLMs, trained on vast datasets, absorb a multitude of signals about norms and values during training. However, while some attention has been given to broad ethical principles like helpfulness and harmlessness, an important aspect remains underexplored: ``contextual rules''\,---\,human norms related to cultural conventions. 
These contextual rules, while not directly influencing primary optimization objectives, are often followed due to tradition, or social norms. 
Despite their indirect nature, such rules can provide valuable signals about broader societal dynamics, thereby guiding the alignment of LLMs, as discussed by \citet{hadfieldmenell2019silly} and~\citet{koster2020silly} within the broader context of AI alignment. Although efforts such as \citet{ziems-etal-2022-moral}, \citet{zhan-etal-2024-renovi}, and \citet{chiu2024culturalteaming} introduce datasets with collections of social norms, the influence of the collected norms on improving alignment in LLMs remains underexplored \citep{aakanksha-etal-2024-multilingual}.
Contextual rules could guide the style of language to align with cultural expectations. For instance, when interacting with users from diverse cultural backgrounds, LLM could account for cultural preferences by avoiding humor that might not translate well across cultures. 
However, existing models have been shown to predominantly reflect Western values, as they have been primarily trained on Western-centric data, which limits their ability to represent multi-cultural values \citep{durmus2024towards,nayak-etal-2024-benchmarking}.\looseness=-1

Human social norms and values are continuously shaped and evaluated through daily interactions with others. These interactions involve the exchange of multimodal signals, such as language, facial expressions, and gestures \citep{doi:10.1098/rstb.2013.0302}. However, when interacting with LLMs, these cues are inherently absent, creating a normative gap in communication. This has prompted research dedicated to auditing values embedded within existing models \citep{huang2025valueswilddiscoveringanalyzing,bhatia2025valuedriftstracingvalue}. 
Further, exploring multimodality for alignment\,---\,integrating non-verbal forms of communication such as visual, or auditory signals\,---\,can serve as a promising line of research to address this normative void. By incorporating multimodal interactions, models could better align with the implicit social expectations typically conveyed through non-verbal cues \citep{liu-etal-2021-visually,nayak-etal-2024-benchmarking}.\looseness=-1

\subsubsection{Allowing for dynamic norms and values} 
Norms and values are not static objects but dynamic equilibria that evolve through ongoing social interactions \citep{doi:10.1073/pnas.1817095116}. They are continuously re-articulated and negotiated within social contexts, evolving to address new challenges and cultural shifts \citep{annurev-psych-033020-013319}. Stereotypes, as a form of social norm, are accordingly fluid, emerging and transforming over time. An example is the shifting perception of remote work. Once seen as unprofessional or less productive, it is now widely accepted in many industries. If an LLM were trained primarily on pre-COVID data, it could reinforce outdated assumptions.

While model editing and continual learning have been extensively explored for updating factual knowledge in LLMs \citep{pmlr-v162-mitchell22a,pmlr-v199-prado22a}, their application for adapting to evolving societal values and norms remains underexplored. Developing approaches to enable LLMs to dynamically identify, adapt to, and mitigate emerging biases is a crucial area for future research. Practically, this involves developing continual learning pipelines that integrate multimodal signals. However, even factual updates pose significant challenges, as highlighted by recent work on knowledge editing \citep{cooper2024machine, hase2024fundamental}.

\subsection{Economic Alignment}
\label{sec:economic}

Economic systems rely on specialization and the division of labor, requiring coordination among groups of people to ensure efficient allocation of resources \citep{arrow1951extension}. A central challenge in modern economic theory is aligning individual actors' interests with collective objectives \citep{hadfieldmanell2019}. Welfare economics provides a complementary perspective by formalizing optimization functions for resource allocation to maximize overall system objectives under given constraints. 
Similarly, aligning LLMs with diverse human values involves navigating trade-offs between individual and collective goals. Additionally, a coherent social welfare objective function for LLMs cannot rely solely on subjective values. Instead, real-world implementations demand collective decisions about which values to prioritize \citep{arrow1951extension,18108}. Building from this, we explore strategies for integrating economic alignment frameworks to coordinate individual preferences to achieve collective, fair objectives and facilitating group-level aggregation, offering an alternative view to imposing monolithic objective functions across diverse user groups.\looseness=-1

\subsubsection{Economic Mechanisms for Fair Alignment} 
In theoretical economics, perfect markets are often posited as achieving a Pareto-efficient distribution of welfare under a utilitarian framework \citep{arrow1951extension}. Pareto efficiency refers to a state where no individual can be made better off without making someone worse off, and is a benchmark for efficient resource allocation \citep{black2017dictionary}. 
In the context of LLMs, this resource allocation translates to the distribution of model performance, representational fairness, or helpfulness across diverse demographic groups or conflicting values.
As shown in \citet{boldi2024paretooptimallearningpreferenceshidden}, Pareto efficiency offers a valuable lens for balancing competing human preferences and optimizing specific notions of group fairness. In theory, achieving such efficiency aims to tailor the model's behavior to address diverse needs. However, we caution that strict Pareto efficiency can inadvertently preserve an inequitable status quo, as an optimized model might favor a majority demographic over a marginalized group simply because helping the minority would slightly degrade the majority's performance. 

To overcome this limitation, LLM alignment can draw on alternative frameworks from social welfare economics, where the aggregation of diverse preferences is explicitly balanced to ensure the true collective well-being of multiple groups, rather than preserving a baseline \citep{DASPREMONT2002459}. Specifically, integrating Social Welfare Functions, such as Nash or Egalitarian welfare objectives, directly into the reward modeling phase mathematically can enforce group-level equity. By operationalizing pluralistic preference aggregation, these objective functions provide a robust guide for the development of reward systems. Indeed, prior work has indicated that developing welfare-centric objectives can enhance fairness outcomes \citep{pardeshi2024learning,cousins2024welfare}.

\subsubsection{Economic Mechanisms for Pluralistic Alignment} 

Decision-making often involves multiple actors with diverse and sometimes conflicting preferences. In the context of LLMs, this necessitates approaches that account for a broad range of values. Pluralistic alignment addresses this challenge by designing models that can represent and respect diverse perspectives \citep{SorensenMFGMRYJ24,tanmay2023probingmoraldevelopmentlarge}. Unlike monolithic approaches, which attempt to impose a singular objective function, pluralistic alignment embraces the complexity of modern societies.\looseness=-1

A critical aspect of LLM alignment involves determining how to elicit and aggregate preferences when multiple humans are affected by the behavior of an artificial agent \citep{rossi2011preferences,rao-etal-2023-ethical,pmlr-v235-conitzer24a}. This challenge extends beyond individual alignment to group alignment, where many societal issues arise from collective behavior rather than isolated actions and can be addressed by incorporating multiple objectives into the alignment process, leveraging methods such as few-shot learning to capture diverse perspectives \citep{zhou-etal-2024-beyond,zhao2024group}.

Another critical issue in enabling pluralistic values is the trade-off between developing general-purpose models and specialized models. While specialized models tailored to specific domains, such as healthcare or justice, can better align with local norms and regulatory frameworks, they risk fragmenting values. Conversely, general-purpose models may provide broader applicability but struggle to adapt to ethically complex, domain-specific requirements. Cooperative game theory offers a framework to navigate these tensions by promoting fair resource allocation, fostering collaboration among stakeholders, and ensuring equitable outcomes \citep{10.5555/2132771}.\looseness=-1

\subsection{Contractual Alignment}
\label{sec:legal}


Law-making and legal interpretation serve as mechanisms to translate opaque human goals and values into explicit, actionable directives. Legal scholars have long recognized the inherent impossibility of drafting complete contracts \citep{macneil1977contracts,williamson1975markets,shavell1980damage,maskin1999unforseen,tirole1999incomplete,aghion2011incomplete}. This limitation stems from several key challenges. First, certain states of the world are either unobservable or unverifiable, e.g., hiding assets in complex financial arrangements can be difficult for tax authorities to identify \citep{69c1e19e-b0f8-307a-abda-571627b432cb}. Second, the limited rationality of humans restricts their ability to anticipate and optimize across the entire, combinatorially large space of potential scenarios \citep{williamson1975markets}. Consequently, precisely computing optimal outcomes becomes intractable. Furthermore, the very description of all possible contingencies is often beyond human foresight, leading to loopholes in the design of rules \citep{183a5147-673b-36d2-8ddd-a7835149772a}. Even if feasible, the costs associated with drafting and enforcing fully specified contracts would likely be prohibitive. 
Given that these challenges are analogous to those encountered in aligning LLMs, where developers aim to ensure that models produce safe and correct outputs even for inputs not directly represented in training or alignment data, we investigate insights from contract theory as potential solutions for improving LLM alignment.

\subsubsection{External Contractual Alignment}
The formalization of contracts offers a framework for anticipating and specifying desired behaviors in human-LLM interactions \citep{jacovi2021formalizing}. 
In this context, standardized documentation plays a crucial role in defining and communicating the LLMs' performance characteristics. Initiatives such as datasheets \citep{10.1145/3458723}, data statements \citep{10.1162/tacl_a_00041}, model cards \citep{10.1145/3287560.3287596}, reproducibility checklists \citep{pineau2020checklist}, fairness checklists \citep{10.1145/3313831.3376445}, and factsheets \citep{8843893} exemplify efforts to create clear, standardized guidelines that could inform the development of future regulations and legal frameworks for LLM alignment and data governance.

The rules that guide LLM alignment are currently largely constructed in consultation with domain and legal experts, by adapting documents such as the UN Declaration of Human Rights \citep{anthropic2023claude}, through public input \citep{anthropic2023collective}, or in some cases, relying on designer instincts \citep{anthropic2023claude,10.5555/3540261.3540709}. 
Importantly, the European Commission has developed detailed guidelines for trustworthy AI, which provide a structured approach to ensuring that AI systems, including LLMs, adhere to ethical principles and societal norms.\footnote{\raggedright The guidelines are available at \url{https://ec.europa.eu/digital-single-market/en/news/ethics-guidelines-trustworthy-ai/}.} 
These documents serve as critical tools for defining the terms of human-LLM contracts and offer a principled way to ensure that the view not only reflects the developer's personal views.

\subsubsection{Internal Contractual Alignment}
While the above discussion focused on aligning LLMs through external rules, another approach takes inspiration from how parties in a contract, laws, and democratic institutions enforce principles. Instead of relying solely on external oversight, this approach embeds normative principles directly within the model's internal mechanisms. Known as \emph{constitutional AI}, this method enables LLMs to develop an internalized set of ``principles'' that guide the model to self-critique and rewrite the response to ensure alignment with predefined norms. By integrating desired rules into the training objectives, constitutional AI aims to instill structural governance within models, much like how legal frameworks encode societal values into enforceable policies.
These methods provide scalable oversight precisely because they move beyond the need for direct, case-by-case human intervention. Traditional preference-based training methods, such as collecting annotations on preferred and rejected outputs, aggregate multiple annotators' judgments into a shared standard, but they still require extensive human effort at scale 
\citep{shen2023largelanguagemodelalignment,amodei2016concreteproblemsaisafety}. 
In contrast, scalable oversight techniques generalize beyond individual preferences by structuring decision-making mechanisms, similar to how democratic systems use institutionalized processes to apply laws across diverse contexts \citep{shen2023largelanguagemodelalignment}. \looseness=-1

One such method, debate \citep{irving2018aisafetydebate,irving2019ai}, mirrors adversarial legal reasoning: agents (i.e., LLMs) propose answers, engage in structured argumentation, and refine their positions, with a human judge selecting the best-supported response \citep{HAFNER20018675}. 
Similarly, constitutional AI guides LLMs using a concise constitution of high-level principles (e.g., promoting fairness or avoiding harm) \citep{bai2022constitutionalaiharmlessnessai,10.5555/3666122.3666237}. This constitution provides the basis for generating synthetic comparison examples, which are then used to fine-tune the LLM's policy. While primarily developed for integrating human values, these methods have the potential to enforce norms and regulations in a structured manner, drawing parallels to how societal governance mechanisms uphold laws and ethical standards.

In practice, as models are increasingly governed by explicit ``constitutions,'' rigorous analytical tools are needed to measure how these rules translate into behavioral changes. For instance, a possible research direction uses statistical distance measures to quantify the distribution shifts in model outputs elicited by new alignment constraints.\looseness=-1

%% file: sections/05_uncertainty.tex
\section{Societal Alignment Frameworks and their View on Uncertainty}
\label{sec:uncertainty}

By framing LLM alignment as a problem of contractual incompleteness and analyzing it through the lens of societal alignment frameworks, we observe that these frameworks recognize establishing contracts, much like alignment, as inherently uncertain \citep{seita1984uncertainty}. 
While the unwanted epistemic uncertainty can undermine the reliability of language models, certain types of uncertainty are not only unavoidable but essential for their ethical deployment \citep{delacroix2024lost}. In the context of LLMs, this essential uncertainty can arise from evolving human values, conflicting societal norms, and the difficulty of translating abstract principles into model behavior.
Aligning models to navigate trade-offs, such as between helpfulness and harmlessness or accuracy and fairness, requires addressing conflicting and often underspecified priorities, which introduces another source of uncertainty \citep{zollo2024prompt, yaghini2023learning}. For instance, when deploying an LLM, we often want to maximize performance subject to some constraints or guardrails on behavior, e.g., a chatbot should give users their desired output, as long as it is not too toxic. The effectiveness of balancing these conflicting priorities and the unintended consequences are often difficult to predict. However, this balancing act is also essential because it allows models to operate within complex, context-dependent environments where rigid adherence to a single objective could lead to harmful outcomes.

Building on the above, the inherent uncertainty in LLM alignment is not a weakness but often a valuable feature that enables models to handle complex scenarios ethically \citep{delacroix2024lost}. To illustrate this, consider an AI system deployed for education, tasked with suggesting new exercises to a student. If the system lacks normative uncertainty, it risks overfitting to a narrow objective, such as strictly minimizing immediate failure rates.  It might continuously propose overly simplistic tasks to a student who initially performed poorly, instead of adapting to the student's cognitive development. By maintaining a degree of normative uncertainty, LLMs can avoid such optimization risks.
This flexibility parallels how legal systems manage the ambiguity of incomplete laws through adaptive mechanisms such as precedent, amendments, and ongoing judicial interpretation. Furthermore, as highlighted by \citet{10.1145/3461702.3462571}, uncertainty communication can be useful for obtaining fairer models by revealing data biases, improving decision-making by guiding reliance on predictions, and building trust in automated systems.

However, effectively communicating this uncertainty remains a challenge. Unlike humans, LLMs lack the non-verbal and contextual cues that naturally support nuanced communication \citep{Bisconti2021}. Existing research has shown that LLMs struggle to convey their uncertainty to users, both implicitly (e.g., through hedging language) and explicitly (e.g., via confidence scores), a skill that humans possess intuitively \citep{Alkaissi2023ArtificialHI, liu2024trustworthyllmssurveyguideline,shorinwa2024surveyuncertaintyquantificationlarge}. 
On the other hand, humans themselves have varying levels of understanding regarding probability and statistics, which are needed to interpret model uncertainty estimates \citep{10.1145/3461702.3462571,10.1001/archinternmed.2009.481}. Human cognition is also subject to inherent biases that can impede accurate interpretation of uncertainty \citep{Kahneman,REYNA200889}. These challenges can be partially addressed by choosing the appropriate communication methods, a key consideration for the design of effective user interfaces \citep{8457476}, and by designing collaborative interaction environments, as discussed by \citet{Montemayor2021}.\looseness=-1

%% file: sections/06_alternative.tex
\section{The Democratic Opportunity Inherent in the Under-specified Nature of LLMs' Objectives}
\label{sec:alternative}

The challenge of aligning LLMs is often framed as a technical problem, one that can be solved through better reward modeling, training objectives, or oversight mechanisms. However, as our exploration suggests, alignment is not merely a technological issue. It is fundamentally a societal one.
To understand the significance of this alternative view, one needs to take a step back and start from the following: we humans are constantly in the process of finding our way around the world. Part of that process involves imagining better ways of living together. We may find some of our practices to be inadequate, for instance, but may not always be able to articulate why. In such cases, we often resort to conversations to refine our intuitions and distill their underlying structures. These evolving dialogues shape and refine our moral and social expectations, which, in turn, influence the values that guide our decision-making. The fact that these values change and often clash is a good sign---a sign of ongoing critical engagement and willingness to question existing norms. 

Now, consider a team of engineers designing AI tools that will be deployed within contexts such as education, healthcare, or justice practices. Some of these tools, like LLMs, can be used as conversational partners. The feedback given as a context can be leveraged to refine LLMs' behavior. Given the inherently dynamic nature of the values that inform education, healthcare, or justice practices, as we previously discussed, the key problem is to establish how to structure this feedback process. Different groups of users will evolve different values over time. Are there ways of incentivizing collective, critical engagement with LLMs? Can bottom-up, iterative refinements be configured to support users in defining the very values that preside over their practices \citep{delacroix2024lost}?\looseness=-1

The ``incomplete contract'' metaphor has been a useful diagnostic tool throughout this paper, highlighting why purely technical attempts to fully specify LLM objectives are bound to fall short, much like real-world contracts cannot anticipate every contingency.
However, we must actively recognize the limitations of this metaphor if taken too literally as a prescriptive framework for solutions. As \citet{goldoni2018} caution, viewing alignment solely as a contract to be \emph{completed} can oversimplify complex systems, potentially framing it as a straightforward, albeit difficult, agreement between clearly defined stakeholders (like a principal and agent). This risks neglecting the broader socio-political forces and conflicting norms that shape alignment challenges, and can inadvertently reinforce a designer-centric, epistemic focus on \emph{fixing} the incompleteness \citep{terzis2024}.
While societal alignment frameworks, as discussed, aim to address these broader issues, they too can rely on oversimplified assumptions if not carefully applied. It is precisely this irreducible uncertainty and the inherent incompleteness of any pre-defined alignment stage that necessitates this alternative view. The true opportunity lies not in striving to technically \emph{perfect} an inherently incomplete contract, but in monitoring how these contracts are formed and by whom. 
The under-specified nature of LLMs' objectives presents a democratic opportunity to democratize the very process of determining what LLMs should optimize for. This reframes the challenge from one of technical specification by a few to one of ongoing, participatory value deliberation by many.

This inherent limitation necessitates an alternative view. As highlighted by recent work on the democratization of AI \citep{Lin_2024}, aligning AI systems requires moving away from a solely technical framing and toward participatory engagement. Furthermore, while recent work by \citet{milliere2025normative} critiques current preference-tuning methods for resulting in ``shallow alignment,'' where models mimic normative patterns without their understanding, our contract theory framework explains \textit{why} this shallowness persists. The true opportunity lies not in striving to technically \emph{perfect} an inherently incomplete contract, but in democratizing the very process of determining what LLMs should optimize for. This reframes the challenge from one of technical specification by a few to one of ongoing, participatory value deliberation by many.

The implications of this reframing extend to both research and practice. Addressing the concentrated power dynamics inherent in the principal-agent relationship requires participatory alignment. This suggests that, alongside technical work such as reward modeling, we need equally sophisticated work on participatory interface designs.
This can be operationalized by integrating iterative feedback loops where affected communities actively shape model behavior, as demonstrated by initiatives like the PRISM dataset \citep{kirk2024}. Rather than relying solely on developer-defined annotations, preference fine-tuning can be conducted on diverse, real-user data inferred from community behaviors and deliberations. This might also include developing new methodologies for collective value articulation \citep{Bergman2024}, creating institutional structures for meaningful public participation in LLM development, and establishing mechanisms for ongoing societal oversight and input into LLMs' objectives and constraints.


%% file: sections/07_related_work.tex
\section{Related Work}

The field of sociotechnical alignment posits that AI systems are embedded within larger institutional structures, including corporations, markets, and nation-states. Consequently, beneficial societal outcomes cannot be guaranteed simply by aligning an individual AI system with the immediate intentions of its operator or user \cite{edelman2025fullstackalignmentcoaligningai, gabriel2020artificial, gabriel2024ethicsadvancedaiassistants, gabriel2025matter}. Sociotechnical alignment aims to address this gap by broadening the scope of AI alignment to incorporate social and systemic factors. In this paper, we specifically focus on the interdisciplinary connections between LLM alignment and established societal frameworks, centering our analysis on the shared challenge of contractual incompleteness.

Since the early works on sociotechnical alignment, a recurring tension persists between ``local'' and ``global'' alignment. Early foundational works have noted that while AI systems may be locally aligned with an operator's specific instructions, they frequently remain misaligned with broader societal interests and objectives \cite{kim2020deep, olson1965logic}. This misalignment often stems from the collective action problems and the inherent difficulty of specifying societal-level welfare in a way that technical systems can optimize. Our work builds upon these insights by framing this underspecification not merely as a technical challenge, but as a fundamental characteristic of incomplete contracting that requires the integration of social, economic, and contractual theories.

A recent development in this area is the concept of \textit{full-stack alignment}, as proposed by \citet{edelman2025fullstackalignmentcoaligningai}, which advocates for the creation of normatively competent agents capable of sophisticated normative reasoning. This mirrors our discussion on the necessity of normative competence for navigating implicit social cues. Furthermore, the challenge of addressing the diverse and often conflicting values of a global user base has led to the emergence of \textit{pluralistic alignment} \cite{SorensenMFGMRYJ24}, which seeks to represent a multitude of perspectives rather than imposing a monolithic standard. From an economic and political perspective, researchers have argued that \textit{social choice theory} should serve as a primary guide for LLM alignment \cite{pmlr-v235-conitzer24a}, providing formal mechanisms for aggregating individual preferences into collective decisions.

%% file: sections/08_conclusion.tex
\section{Conclusions}

Aligning LLMs with human values remains a critical challenge as these models are increasingly deployed in real-world applications. Current approaches, such as RLHF, struggle with the inherent misspecification of reward functions, which fail to capture the complexity and evolving nature of human values. Additionally, language ambiguity further complicates alignment efforts.\looseness=-1 

In this paper, we have argued that addressing these challenges requires reframing LLM alignment through the lens of contract theory. We model LLM alignment within a principal-agent framework, where the principal (a user or developer) defines a contract. This contract consists of an action taken by the agent (the LLM) and the corresponding reward assigned by the principal. We then draw connections between the challenges of contract formation in societal alignment frameworks and those in LLM alignment, arguing that insights from societal alignment can improve LLM alignment within incomplete contracting environments. While contract theory provides us with some formalization tools, social alignment emphasizes the role of instillation of societal norms and values, economic alignment points to solutions to group alignment and allocation challenges, and contractual alignment provides mechanisms for regulating LLM behavior, both externally through legal frameworks and internally through scalable oversight mechanisms. Finally, we present an alternative view on LLM alignment, advocating for shifting the paradigm from developer-centered to collaborative, user-centric, and iterative approaches to LLM alignment.\looseness=-1
